\documentclass[aps,12pt,preprint,superscriptaddress,nofootinbib]{revtex4-2}
\usepackage[utf8]{inputenc}
\usepackage{amsthm,amsmath,physics,mathtools,amssymb}

\usepackage{charter}

\usepackage{CJKutf8}

\usepackage{enumitem}

\usepackage{xcolor}
\usepackage{hyperref}
\hypersetup{
    colorlinks=true,
    citecolor=blue,
    linkcolor=blue,
    filecolor=magenta,      
    urlcolor=cyan,
}
\usepackage[capitalise]{cleveref}

\theoremstyle{definition}

\usepackage{graphicx}
\graphicspath{{figures/}}

\usepackage{tensor}

\DeclarePairedDelimiterX{\inp}[2]{\langle}{\rangle}{#1, #2}

\usepackage{xparse}

\NewDocumentCommand\LH{mo}{%
  \IfNoValueTF{#2}
   {\mathcal{L}(\mathcal{H}^{#1})}
   {\mathcal{L}(\mathcal{H}^{#1},\mathcal{H}^{#2})}%
}

\newcommand\id{\leavevmode\hbox{\small1\kern-3.3pt\normalsize1}}

\allowdisplaybreaks

\usepackage{bbold}

\usepackage[title]{appendix}

\usepackage{mciteplus}

\usepackage{enumitem} 

\usepackage{float}

\begin{document}
\title{Truly Lorentzian quantum cosmology}
\begin{CJK*}{UTF8}{gbsn}
\author{Ding Jia (贾丁)}
\email{djia@perimeterinstitute.ca}
\affiliation{Perimeter Institute for Theoretical Physics, Waterloo, Ontario, N2L 2Y5, Canada}
\affiliation{Department of Applied Mathematics, University of Waterloo, Waterloo, Ontario, N2L 3G1, Canada}

\begin{abstract}
Quantum cosmology based on Lorentzian path integrals is a promising avenue. However, many previous works allow non-Lorentzian configurations by integrating the squared scale factor over the whole real line. Here we show that restricting the minisuperspace path integral to Lorentzian configurations with positive squared scale factor can significantly change the expectation values. In addition, this enables the study of causal horizons and their quantum fluctuations, and achieves singularity avoidance trivially by excluding singular minisuperspace geometries as non-Lorentzian. The results indicate that semiclassical saddle point approximation is not always valid in truly Lorentzian quantum cosmology. As a consequence, related works on the tunnelling and no-boundary proposals, bouncing cosmology, and the quantum origin of inflation etc. need to be reexamined.
\end{abstract}

\maketitle
\newpage
\end{CJK*}

\section{Introduction}\label{sec:i}

Transitioning to the Lorentzian signature has been a recurring theme in quantum gravity. Historically, approaches such as simplicial quantum gravity \cite{Hamber2009QuantumApproach}, dynamical triangulation \cite{AmbjornQuantizationGeometry}, spin-foam and related gauge theories \cite{Regge2000DiscreteGravity} started in the Euclidean. Subsequent works eventually encompassed the Lorentzian setting to counter issues such as spikes \cite{Tate2011Fixed-topologyDomain, Jia2022Time-spaceGravity}, degenerate geometries \cite{Ambjrn1998Non-perturbativeChange, Loll2003AIntegral}, conformal instabilities \cite{Smolin1979QuantumLattice}, or to simply engage with \textit{spacetime} which is Lorentzian.

In quantum cosmology one studies simplified models of quantum gravity such as the homogeneous and isotropic minisuperspace model with the metric
\begin{align} \label{eq:metric1}
	\dd s^2=-N^2\dd \tau^2+a(\tau)^2\dd\Omega^2,
\end{align}
where $\dd\Omega^2$ is the metric of a closed spatial 3-sphere.
Since Euclidean gravitational path integrals suffer from the conformal instability issue \cite{Gibbons1977TheThermodynamics}, old works explored different complex integration contours \cite{Hartle1985SimplicialDiscussion, Hartle1986SimplicialTriangulations, Hartle1989SimplicalModel, Halliwell1989Steepest-descentModel, Halliwell1989Steepest-descentMicrosuperspace, Halliwell1990Steepest-descentModels, Halliwell1989Multiple-sphereUniverse, Halliwell1990IntegrationUniverse, Schleich1985SemiclassicalThree-geometries, Schleich1989ConformalCosmology}. There have been various discussions about fixing the integration contour to be over Lorentzian spacetimes in the past (e.g., \cite{Farhi1989TheOne, Cline1989DoesGravity, Suen1989WaveSystem, Brown1990LorentzianCosmology}). More recently, Feldbrugge \textit{et al.} proposed \cite{Feldbrugge2017LorentzianCosmology} to define the gravitational path integrals by the Lorentzian contour, and use Picard-Lefschetz theory to study complex contour deformations only as a computational trick for the fundamentally Lorentzian theory (see Sorkin \cite{Sorkin2013IsLorentzian} for a closely related discussion). 

This has led to renewed interest in investigating old topics with new methods \cite{Feldbrugge2017LorentzianCosmology, Dorronsoro2017RealCosmology, Feldbrugge2017NoSpacetime, Feldbrugge2017NoProposal, Dorronsoro2018DampedState, Feldbrugge2018InconsistenciesProposal,  Vilenkin2018TunnelingUniverse, DiTucci2018UnstableMetrics, DiTucci2019QuantumInflation, Janssen2019TheMinisuperspace, DiTucci2019No-BoundaryConditions, DiTucci2019No-boundaryCosmology, Bramberger2019HomogeneousCosmology, DiTucci2020LessonsHoles, NarainLorentzianMethods, Rajeev2021No-boundaryCosmology, Rajeev2021BouncingCosmology, Lehners2021ThePotentials, Narain2021OnQuantization, Lehners2022AllowableCosmology, Jonas2022RevisitingField, Matsui2022LorentzianWave-function, Narain2022SurprisesGravity, Jonas2022UsesCosmology}.
In these works of ``Lorentzian quantum cosmology '', it is common to adopt the minisuperspace metric
\begin{align} \label{eq:metric2}
	\dd s^2=-\frac{N^2}{q(t)}\dd t^2+q(t)\dd\Omega^2,
\end{align}
and treat the squared scale factor $q$ as a path integral variable. The action in $q$ is quadratic and a Gaussian integration yields nice closed-form results \cite{Halliwell1988DerivationModels, Halliwell1989Steepest-descentModel}. 

However, the Gaussian integration over all real values of $q$ is questionable step. For the metric to stay in the Lorentzian signature, $q$ should only assume positive values. A path integral over also negative $q$ cannot be said to be truly Lorentzian. In the seminal paper \cite{Halliwell1988DerivationModels} that adopted $q$ as a path integration variable, Halliwell warned us that
\begin{quote}
In Sec. VII we encountered the problem of doing quantum mechanics in terms of the variable $q$ whose physical range was the positive real line. This problem is not in any way an artifact of the particular model under consideration, but is a manifestation of the fact that the three-metric $h_{ij}$ satisfies the condition $\det h_{ij}>0$. It is therefore important to face up to this issue from the very beginning.
\end{quote}
Yet this warning is largely left aside in subsequent works.

In this work we consider quantum cosmology in a truly Lorentzian setting, where the path integral is only over strictly Lorentzian configurations. In particular, we study minisuperspace quantum cosmology based on the metric \eqref{eq:metric2}, and distinguish the \textbf{real $q$ scheme}, where $q$ is integrated over the whole real line, from the \textbf{positive $q$ scheme}, where $q$ is integrated over positive values. 

We find that the two schemes differ in at least three important aspects. First, the expectation values for the squared scale factor can differ much in the two schemes, when $q$ gets close to or below zero for a relevant saddle point of the path integral. This affects the studies of tunnelling and no-boundary proposals, bouncing cosmology, and the quantum origin of inflation. Second, it is only possible to study the causal horizons and their quantum fluctuations in the positive $q$ scheme. In the real $q$ scheme the path integral includes non-Lorentzian geometries, where causal horizon is not defined. Third, restricting the path integral to the Lorentzian implies singularity avoidance. This is because singular minisuperspace geometries are non-Lorentzian and hence are automatically excluded from the path integral. In this sense, singularity avoidance is trivially achieved \cite{JiaIsTrivial} in the truly Lorentzian minisuperspace path integral. 

The results challenge the universal validity of semiclassical saddle point approximation. In particular, for negative spatial curvature bouncing cosmology, our numerical results based on the generalized thimble method \cite{Alexandru2016SignThimbles} show that the saddle point which dominates the real $q$ scheme path integral completely fails to capture the quantum expectation values of the truly Lorentzian positive $q$ scheme. The true expectation values rather resembles that of the zero spatial curvature case in their real parts, in addition to possessing a large imaginary part. Here neither real nor complex (tunnelling) solutions to Einstein's equations characterize the path integral at leading order, because neither does the saddle point belong to the Lorentzian integration contour, nor does it connect to any configuration of the Lorentzian contour through the Picard-Lefschetz holomorphic gradient flow. As a consequence, semiclassical saddle point approximation should only be applied when its validity can be ascertained. 

The paper is organized as follows. In \Cref{sec:pw} we review recent works on Lorentzian quantum cosmology. In \Cref{sec:lrqs} we point out the limitations of the real $q$ scheme. In \Cref{sec:gtm} we review the generalized thimble method which we use for numerical computation. In \Cref{sec:lf} we define the quantities of lightcone location and its fluctuations to be computed. In \Cref{sec:csbc} we put the pieces together to study bouncing cosmology for positive, zero, and negative spatial curvatures, and compare results from the positive and real $q$ schemes. In \Cref{sec:sa} we discuss singularity avoidance. In \Cref{sec:d} we conclude with a discussion of some topics for further study.

In the following, we set $c=\hbar=8\pi G=1$.

\section{Previous works}\label{sec:pw}

Following \cite{Feldbrugge2017LorentzianCosmology, Halliwell1988DerivationModels, Halliwell1989Steepest-descentModel} we consider the minisuperspace metric
\begin{align} \label{eq:metric}
	\dd s^2=-\frac{N^2}{q(t)}\dd t^2+q(t)\left(\frac{1}{1-kr^2}\dd r^2+r^2\left(\dd\theta^2+\sin^2\theta \dd\phi^2\right)\right),
\end{align}
with squared scale factor $q(t)$, lapse $N$, and spatial curvature $k=1,0$ or $-1$.\footnote{For the flat ($k=0$) and hyperbolic ($k=-1$) cases, we assume that the spatial geometry is compactified as in \cite{Halliwell1990Steepest-descentModels} so that the action \eqref{eq:action} is finite.} The $dN/dt=0$ gauge is used so $N$ does not depend on $t$. 

Plugging \eqref{eq:metric} in the Einstein-Hilbert action with the Gibbons-Hawking-York boundary term \cite{York1972RoleGravitation, Gibbons1977ActionGravity} $S = \frac{1}{2}\int \dd^4x \sqrt{-g} ( R - 2 \Lambda) + \int_B \dd^3y \sqrt{h} K$, one obtains
\begin{align}\label{eq:action}
S[q,N]=2\pi^2\int_0^1\dd t\,N\left(-\frac{3\dot{q}^2}{4N^2}-\Lambda q+3k\right).
\end{align}
The dot denotes derivative with respect to the coordinate time $t$, which is taken to run from $0$ to $1$. This is without loss of generality, since the physical proper time derived from \eqref{eq:metric} is still arbitrary due to $N$ and $q$. 

For the boundary condition $q(0)=q_0, q(1)=q_1$ the path integral takes the form
\begin{align}\label{eq:Z}
Z[q_0,q_1]=\int DN \int_{q_0,q_1} Dq~ e^{iS[q,N]}.
\end{align}
We omit the subscript $q_0,q_1$ below when no ambiguity arises.

The metric \eqref{eq:metric} is written in terms of the squared scale factor rather than the scale factor $a(t)$. This produces the action \eqref{eq:action} which is quadratic in $q$. In many previous works, such as 
\cite{Feldbrugge2017LorentzianCosmology, Dorronsoro2017RealCosmology, Feldbrugge2017NoSpacetime, Feldbrugge2017NoProposal, Dorronsoro2018DampedState, Feldbrugge2018InconsistenciesProposal,  Vilenkin2018TunnelingUniverse, DiTucci2018UnstableMetrics, DiTucci2019QuantumInflation, Janssen2019TheMinisuperspace, DiTucci2019No-BoundaryConditions, DiTucci2019No-boundaryCosmology, Bramberger2019HomogeneousCosmology, DiTucci2020LessonsHoles, NarainLorentzianMethods, Rajeev2021No-boundaryCosmology, Rajeev2021BouncingCosmology, Lehners2021ThePotentials, Narain2021OnQuantization, Lehners2022AllowableCosmology, Jonas2022RevisitingField, Matsui2022LorentzianWave-function, Narain2022SurprisesGravity, Jonas2022UsesCosmology}, the integration range of the squared scale factor is \textit{taken to be over the whole real line}. Then the path integral in $q$ with the quadratic action can be evaluated analytically, just like the path integral of a free quantum particle \cite{Feynman1965QuantumIntegrals}.

Explicitly, a Gaussian functional integration with respect to $q$ yields
\begin{align}\label{eq:prop}
G[q_0,q_1;N]:=&\int Dq~ e^{iS[q,N]}=\sqrt{\frac{3\pi i}{2N}} e^{i 2\pi^2 \int_0^1 \mathrm{d}t ( -\frac{3}{4 N}\dot{\bar{q}}^2 +N (3k- \Lambda \bar{q}) )},
\\
\bar{q}(t)=&\frac{\Lambda}{3}N^2 t^2 + \left(- \frac{\Lambda}{3}N^2+ q_1- q_0\right) t + q_0.\label{eq:qbar}
\end{align}
Here $\bar{q}(t)$ obeys the boundary condition $q(0)=q_0, q(1)=q_1$ and solves
\begin{align}\label{eq:q} 
\ddot{q}  =&  \frac{2\Lambda}{3}N^2,
\end{align}
which is the equation of motion obtained from $\delta S/\delta q=0$.

To obtain the final result $Z[q_0,q_1]=\int DN ~G[q_0,q_1;N]$ from \eqref{eq:prop}, one still needs to analyze the $N$ integral. In previous works this one-dimensional integral is commonly studied through a saddle point approximation. The saddle point can be obtained directly by demanding stationary phase for $G[q_0,q_1;N]$. Since the phase $2\pi^2  \int_0^1 \mathrm{d}t ( -\frac{3}{4 N}\dot{\bar{q}}^2 +N (3k- \Lambda \bar{q}) )$ equals $S[\bar{q},N]$, we have $\partial_N S[\bar{q},N]=0$. Equivalently, we could look at the original path integral \eqref{eq:Z} and demand $\delta S[q,N]/\delta N=0$ to obtain the equation of motion
\begin{align}\label{eq:N} 
\int_0^1 \dd t \left( \frac{3}{4 N^2} \dot{q}^2 +3k-\Lambda q \right)=0.
\end{align}
\Cref{eq:N} and \eqref{eq:q} form the complete set of equations of motion for the variables $q$ and $N$. In the joint solution, $q$ is given by \eqref{eq:qbar}, and $N$ is given by
\begin{align}\label{eq:Nbar} 
\bar{N}=c_1 \frac{3}{\Lambda} \left( ( \frac{\Lambda}{3} q_0-k)^{1/2} + c_2 ( \frac{\Lambda}{3} q_1-k)^{1/2}\right),
\end{align}
where $c_1,c_2\in\{-1,1\}$.

This offers four possible saddle points. However, not all of them will make contributions to the Lorentzian path integral, and Picard-Lefschetz theory can be employed to determine the actually relevant saddle points \cite{Feldbrugge2017LorentzianCosmology}. Previous works show that the relevancy of the saddle points depends on whether the $N$-integral is defined as $\int_{0}^\infty \dd N$ or $\int_{-\infty}^\infty \dd N$. There has been no consensus in the literature on which measure to use \cite{Feldbrugge2017LorentzianCosmology, Dorronsoro2017RealCosmology, Feldbrugge2017NoSpacetime, Feldbrugge2017NoProposal, Dorronsoro2018DampedState, Feldbrugge2018InconsistenciesProposal}.

\section{Limitations of the real q scheme}\label{sec:lrqs}

\subsection{Cases with limitations}\label{sec:cwl}

In the procedure reviewed above, integrating $q(t)$ over the whole real line is crucial. It enables Gaussian integration to obtain the analytic result \eqref{eq:prop}.

However, in the context of \textit{Lorentzian} quantum cosmology there is an unsettling issue. For the metric \eqref{eq:metric} to stay in the Lorentzian signature $(-,+,+,+)$, it must be that $q(t)>0$. Therefore $q$ should only be integrated over positive values in a strictly Lorentzian path integral. 

In practice, integrating $q$ over the real line could still be employed as a useful trick if the result agrees well with integrating over positive $q$. For instance, if the saddle point $\bar{q}$ of \eqref{eq:qbar} stays far above zero for the whole time $t\in[0,1]$, then the integrals in both positive and real $q$ schemes are dominated by paths which stay positive.

\begin{figure}[H]
    \centering
    \includegraphics[width=1.0\textwidth]{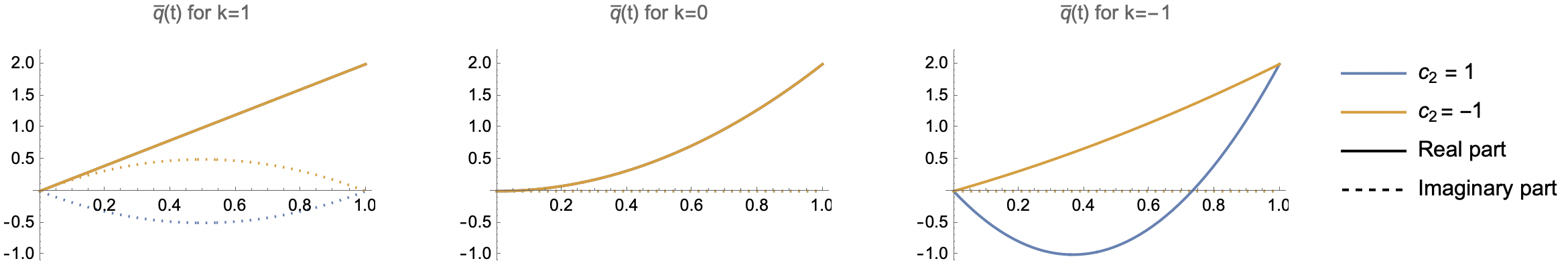}
    \caption{Plotting $\bar{q}(t)$ with $\Lambda=3, q_0=0, q_1=2$ for on-shell $N$ with $c_1=1$. In the first two cases $\Re \bar{q}$ overlap for $c_2=1$ and $c_2=-1$.}
    \label{fig:q-cosmology-cases_1}
\end{figure} 

\begin{figure}[H]
    \centering
    \includegraphics[width=1.0\textwidth]{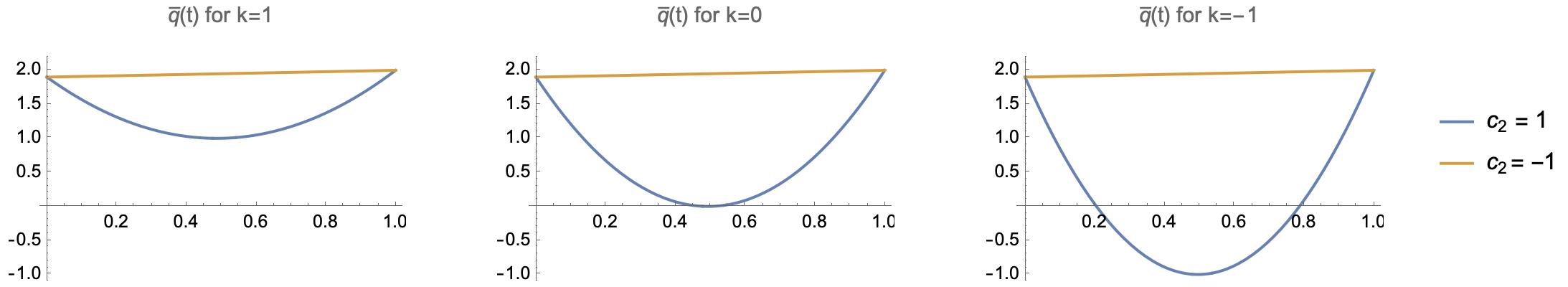}
    \caption{Plotting $\bar{q}(t)$ with $\Lambda=3, q_0=1.9, q_1=2$ for on-shell $N$ with $c_1=1$.}
    \label{fig:q-cosmology-cases_2}
\end{figure}

Yet in some cases $\bar{q}$ does not stay far above zero:
\begin{itemize}
\item When a boundary value $q_0$ or $q_1$ is close to zero, then \eqref{eq:qbar} clearly does not stay far away from zero for all time (\Cref{fig:q-cosmology-cases_1}). This happens for Lorentzian versions (e.g. \cite{Feldbrugge2017LorentzianCosmology, Dorronsoro2017RealCosmology, Vilenkin2018TunnelingUniverse, DiTucci2019No-BoundaryConditions, DiTucci2019No-boundaryCosmology}) of the tunnelling \cite{Vilenkin1982CreationNothing, Vilenkin1984QuantumUniverses, Linde1984QuantumUniverse} and no-boundary \cite{Hartle1983WaveUniverse} boundary conditions where $q_0$ is sent to zero.
\item When $N$ is on-shell at \eqref{eq:Nbar}, the saddle point $\bar{q}$ can reach $3k/\Lambda$ (e.g., at the minimum value for the $c_2=1$ cases in \Cref{fig:q-cosmology-cases_2}). This minimum value can get close to or below zero. For example, this happens in the $k=0$ case relevant to inflation \cite{DiTucci2019QuantumInflation}.
\item When $N$ is allowed off-shell, there are more cases where $\bar{q}(t)$ gets close to or below zero. For example, when $\Lambda>0$ the bouncing saddle point $\bar{q}$ always dives into negative values for sufficiently large $N$.\footnote{The general solution \eqref{eq:qbar} is a parabola with axis of symmetry at $t_a=1/2 + 3(q_0 - q_1)/(2 N^2 \Lambda)$. When $\Lambda>0$, $\bar{q}$ assumes its minimum value
\begin{align}
\bar{q}(t_a)=\frac{1}{12} \left(-\Lambda  N^2-\frac{9 (q_0-q_1)^2}{\Lambda  N^2}+6 (q_0+q_1)\right),
\end{align}
which is always negative for large enough $N$. Moreover, $t_a$ approaches $1/2$ for large $N$, so it always fall within the relevant range $t\in [0,1]$.}
\end{itemize}
The first two cases are especially troublesome. Here the relevant saddle points with both $q$ and $N$ set on-shell get close to or below zero. Paths at and around these saddle points make significant contributions to the path integral in the real $q$ scheme, but are excluded in the positive $q$ scheme. Therefore the real $q$ scheme result may deviate much from the truly Lorentzian positive $q$ scheme result.
\begin{itemize}
\item In addition, for any values of $q_0, q_1, \Lambda, k$, path integral configurations with zero or negative $q$ at some time are not Lorentzian so do not possess a causal structure.
\end{itemize}
This poses a difficulty in studying topics related to causal structures, for example, on the topics of how quantum fluctuations of spacetime affects the horizon problem \cite{Jia2022LightGravity} and light/gravitational wave propagations for bouncing cosmology \cite{Barrau2017SeeingModels}.

\subsection{A toy model example}\label{sec:tme}

To illustrate how the positive and real $q$ schemes can produce quantitatively very different results, we look at a simple toy model just for the $q$ path integral. The results of the later sections, which demand more efforts to obtain, will show that the same happens for the joint $q-N$ path integral. The $q$ path integral can be approximated by
\begin{align}\label{eq:ZN}
G[q_0,q_1;N]\approx&\int Dq ~ e^{i\sum_{i=0}^{n} S_i},
\\
\int Dq =& \int_0^\infty dq(t_1)\cdots \int_0^\infty dq(t_{n}) ~ \mu\left(q(t_1),\cdots, q(t_{n}),N\right),
\label{eq:im}
\\
S_i=&2 \pi^2 \left(-\frac{3\left(q(t_{i+1})-q(t_{i})\right)^2}{4N \Delta t}+N \Delta t \left(3k-\frac{1}{2} \Lambda  \left(q(t_{i})+q(t_{i+1})\right)\right)\right),
\end{align}
where the time domain $t\in [0,1]$ is broken into $n+1$ intervals of size $\Delta t=1/(n+1)$ with the actions $S_i$. The exact result is approached as $n \rightarrow \infty$. 

Here $\mu$ is the measure factor for the integrals. The result \eqref{eq:prop} is obtained with
\begin{align}\label{eq:qm}
\mu=\left(\frac{3\pi i}{2N\Delta t}\right)^{\frac{n+1}{2}},
\end{align} 
which takes the same form of the measure factor for a quantum particle \cite{Feynman1965QuantumIntegrals}. We will use this measure to make the comparison between the positive and real $q$ schemes.


\begin{figure}
    \centering
    \includegraphics[width=.5\textwidth]{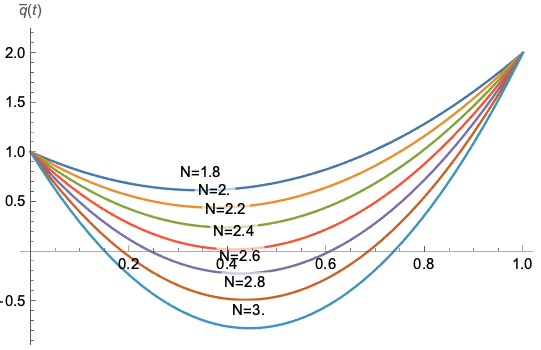}
    \caption{Plotting $\bar{q}(t)$ with $\Lambda=3, k=1, q_0=1, q_1=2$ for a list of $N$ values.}
    \label{fig:plot_qbar}
\end{figure} 
\begin{figure}
    \centering
    \includegraphics[width=1.\textwidth]{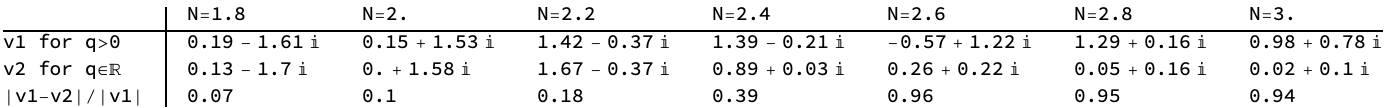}
    \caption{Numerical integration results for the $n=1$ approximation. The first row shows results of the positive $q$ scheme, the second row shows results of the real $q$ scheme, and the third row shows their relative differences.}
    \label{fig:tb_intS2}
\end{figure} 

\Cref{fig:plot_qbar} plots the on-shell $\bar{q}(t)$ of \eqref{eq:qbar} for $\Lambda=3, k=1, q_0=1, q_1=2$ for a list of $N$ values, including some for which $\bar{q}(t)$ gets close to zero or reach negative values. In the simplest approximation $n=1$, there is only one dynamical variable $q:=q(t_1)$, and \eqref{eq:ZN} can be computed by direct numerical integration. The results obtained using \textit{Mathematica} for the positive and real $q$ schemes are shown in \Cref{fig:tb_intS2}. 
Clearly as $N$ increases and $\bar{q}(t)$ approaches zero or negative values, the difference becomes quite significant. Already at $N=2.4$ where $\bar{q}(t)$ still stays positive, the difference reaches as high as $39\%$.

\section{Generalized thimble method}\label{sec:gtm}

In order to investigate the differences between the truly Lorentzian positive $q$ scheme and the real $q$ scheme further, we need a method to evaluate the truly Lorentzian path integrals. The problem is quite non-trivial because analytically, not much is known for path integral computations beyond Gaussian integration. Even numerically, the complex Lorentzian path integral has an oscillating phase that gives rise to the numerical sign problem. 

\subsection{Review of the method}\label{sec:rm}

Fortunately, the generalized thimble numerical method \cite{Alexandru2016SignThimbles, Alexandru2017MonteCarloModel} offers a way to overcome the sign problem. This is a Monte Carlo sampling method that exploits Picard-Lefschetz theory to deform the integration contour to reduce the complex phase fluctuations. It can be viewed as a generalization of the Lefschetz thimble method \cite{AuroraScienceCollaboration2012HighThimble} to other than the steepest descent contours \cite{Alexandru2022ComplexProblem}, which makes the method more adaptable to attack problems such as multimodal problems \cite{Alexandru2017TemperedThimbles, Fukuma2017ParallelThimbles, Fukuma2021WorldvolumeMethod, Fujisawa2022BackpropagatingCalculations}. 

\begin{figure}
    \centering
    \includegraphics[width=.4\textwidth]{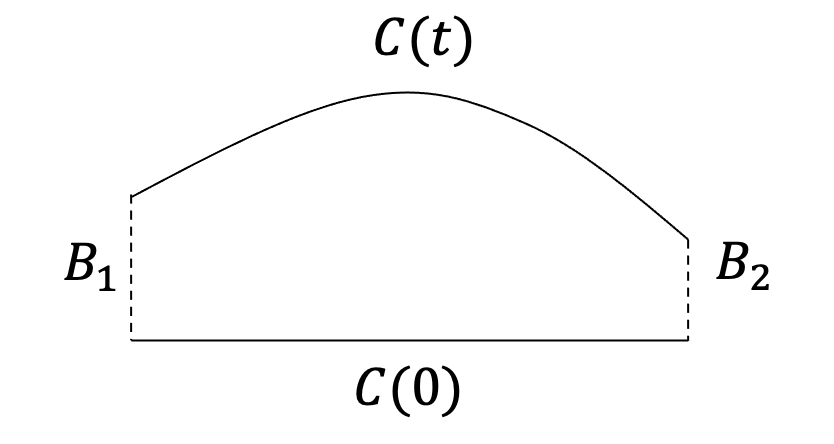}
    \caption{If the integrand is holomorphic in the region enclosed by the curves shown, the integral along the boundary will vanish by Cauchy's integration theorem. As a consequence, the integrals along the contours $C(0)$ and $C(t)$ differ only by the integrals along the dashed boundaries.}
    \label{fig:complex-contour}
\end{figure}

Given a multidimensional integral
\begin{align}\label{eq:piE}
\int \prod_i dv_i ~ e^{E[v_1,v_2,\cdots]},
\end{align}
the \textbf{holomorphic gradient flow} equations
\begin{align}\label{eq:fe}
\frac{d v_i}{dt}=&-\overline{\pdv{E}{v_i}} \quad \forall i
\end{align}
generates an integral curve for each point $\zeta=(v_1,v_2,\cdots,v_n)$ in the original integration contour. Subjecting the whole integration contour to this flow generates a contour deformation $C(t)$ as a function of the flow time $t$, with $C(0)$ as the original contour. If the integrand is holomorphic everywhere the flow transverses, Cauchy's integration theorem guarantees that the integral along $C(t)$ differs from the original one only along the boundaries of the flowed region (\Cref{fig:complex-contour}). If the boundary contributions are negligible, we could use the integral along $C(t)$ to approximate the original integral.

Evaluating the integral along $C(t)$
could ameliorate the sign problem. To see this, note that by \eqref{eq:fe} the real part $E_R$ of $E$ obeys
\begin{align}\label{eq:DreEDt}
\dv{E_R}{t}=&\frac{1}{2}(\dv{E}{t}+\overline{\dv{E}{t}})=\frac{1}{2}\sum_i (\pdv{E}{v_i} \dv{v_i}{t}+\overline{\pdv{E}{v_i} \dv{v_i}{t}})=-\sum_i\abs{\pdv{E}{v_i}}^2\le 0.
\end{align}
Therefore the magnitude of the integrand is exponentially suppressed along the flow, except for regions close to the stationary points where $\pdv*{E}{v_i}=0, \forall i$. For sufficiently large $t$, only this region contributes significantly to the integral along $C(t)$, and we can hope that the phase fluctuation of the integrand is milder than over the original contour.

The generalized thimble method of \cite{Alexandru2016SignThimbles, Alexandru2017MonteCarloModel} exploits the deformed contour to perform Markov Chain Monte Carlo sampling based on the following algorithm:
\begin{enumerate}
\item Fix some flow time $t=T$. Start with a configuration $\zeta=\{v_i\}_i$ in the original contour. Use numerical integration to evolve it under \eqref{eq:fe} by $T$ to obtain $\phi=\phi(\zeta)$.
\item Sample a new configuration $\zeta'=\zeta+\delta\zeta$ on the original contour and evolve $\zeta'$ under \eqref{eq:fe} again by $T$ to obtain $\phi'=\phi'(\zeta')$.
\item Accept $\zeta'$ as the new $\zeta$ with probability $P = \min\{1, e^{\Re E_{\text{eff}}(\phi')-\Re E_{\text{eff}}(\phi)} \}$, where $E_{\text{eff}}$ is defined below in \eqref{eq:Eeff}.
\item Repeat steps 2 and 3 until a sufficient ensemble of configurations is generated.
\item Compute the expectation values using the formula
\begin{align}\label{eq:expo}
\ev{O}=&\frac{\ev{O e^{i\varphi(\zeta)}}_{\Re E_{\text{eff}}}}{\ev{e^{i\varphi(\zeta)}}_{\Re E_{\text{eff}}}},
\end{align}
where $\varphi$ is defined in (\ref{eq:vphi}) and $\ev{\cdot}_{\Re E_{\text{eff}}}$ denotes averaging over the ensemble just generated.
\end{enumerate}
\noindent To define $E_{\text{eff}}$, we note that
\begin{align}
\int_{C(0)} e^{E(\zeta)} d\zeta = \int_{C(T)} e^{E(\phi)} d\phi=\int_{C(0)} e^{E(\phi(\zeta))} \det J ~d\zeta.
\label{eq:ppi}
\end{align}
In the second expression, the contour $C(T)$ is parametrized by the flowed coordinates $\phi$. In the last expression the contour $C(T)$ is reparametrized by the original coordinates $\zeta$. This induces the Jacobian $J_{ij} = \pdv{\phi_i}{\zeta_{j}}$ which can be computed by integrating to $t=T$
\begin{align}\label{eq:jcb}
\frac{d J_{ij}(t)}{dt}= \sum_{k}\overline{H_{ik}J_{kj}},\quad H_{ij}=-\pdv{E}{v_i}{v_j}, \quad J_{ij}(0)=\delta_{ij}.
\end{align}
The integrand exponent of the last integral of \eqref{eq:ppi} is given a special name
\begin{align}\label{eq:Eeff}
    E_{\text{eff}}=E(\phi(\zeta)) + \log \det J(\zeta)
\end{align}
Expanding $E_{\text{eff}}$ in real and imaginary parts yields $e^{E_{\text{eff}}}= e^{\Re E_{\text{eff}}  + i \varphi}$, where
\begin{align}\label{eq:vphi}
\varphi= \Im E_{\text{eff}}=\Im E + \arg\det(J).
\end{align}
This explains steps 3 and 5, in which we sample (\ref{eq:ppi}) according to the magnitude $e^{\Re E_{\text{eff}}}$ of the integrand, and multiply $O$ with the phase $e^{i\varphi}$ in (\ref{eq:expo}).

\subsection{Integration range and measure factors}

We want to apply the generalized thimble method to the path integral \eqref{eq:Z}
\begin{align}
Z[q_0,q_1]=\int DN \int Dq~ e^{iS[q,N]}=\int DN ~G[q_0,q_1;N]
\end{align}
with $G[q_0,q_1;N]$ given in \eqref{eq:ZN}. For this we need to specify the integration range and measure factors.

Since the metric \eqref{eq:metric} is of the Lorentzian signature $(-,+,+,+)$ only when $q$ is positive, we integrate $q$ over positive values as in \eqref{eq:im}. 

For the $q$ measure factor $\mu$ of \eqref{eq:im}, previous results in the real $q$ scheme employed \eqref{eq:qm}. Since $\int_{-\infty}^\infty e^{-ax^2} dx=\sqrt{\pi/a}=2\int_{0}^\infty e^{-ax^2} dx$ for $\Re a>0$, it seems reasonable to modify \eqref{eq:qm} by a constant factor in the positive $q$ scheme. Since constant multiplicative factors cancel out in \eqref{eq:expo}, for simplicity we will directly employ $\mu=(\frac{1}{N})^{\frac{n+1}{2}}$. This factor $\mu$ can be incorporated as an additional term
\begin{align}\label{eq:qm2}
E_\mu=\frac{n+1}{2}\log N
\end{align}
in the path integral exponent $E$ of \eqref{eq:piE}.

As mentioned at the end of \Cref{sec:pw}, there is more than one choice for the $N$ integration range. Here we take
\begin{align}\label{eq:Nm}
\int DN=\int_0^\infty \dd N
\end{align}
as in \cite{Feldbrugge2017LorentzianCosmology} for concreteness.
The disagreement between real and positive $q$ schemes should also be present for the alternative measure $\int_{-\infty}^\infty \dd N$. Although we have not performed the study, it seems the generalized thimble method can be applied to this case as well.

An additional measure factor is included for the following reason. The $q$ and $N$ integration ranges are both bounded by $0$. As explained around \Cref{fig:complex-contour}, the integrals along the flowed contour and the original contour agree well if contributions are small along the ``side contours'' (dashed part of \Cref{fig:complex-contour}) traced by the boundaries of the original contour under the holomorphic gradient flow. For the particular case at hand, the boundaries of the original lie at $N=0$ and $q(t_i)=0, ~ \forall i$. According to \eqref{eq:qm2}, the $N=0$ boundary is a log singularity.
As explained in Section 2 of \cite{Feldbrugge2017LorentzianCosmology}, such a singularity is unchanged under the holomorphic gradient flow. As the flow time grows from $T=0$ to $T=\infty$ toward the Lefschetz thimbles, the contour is pinched at $N=0$ while the angle at which the contour reaches $N=0$ varies. Therefore this ``side contour'' has no extension, and offers no contribution to the integral. 

To ensure that ``side contour'' also offers no contribution to the integral at the $q(t_i)=0$ boundaries, we employ the trick of introducing the measure factor $\prod_i q(t_i)^m$ for some $m<0$. For $\abs{m}\ll 1$ such as $m=-0.001$ used here, $q^m$ stays fairly close to one for practical ranges of $q$. For instance, consider the range starting from the very tiny value of $q=10^{-8}$ and extending to the very large value of $q=10^{2}$ in comparison to the saddle paths of \Cref{sec:csbc}. In this range, $q^m$ only decreases monotonically from $1.0186$ to $0.9954$ to four digits after the decimal place for $m=-0.001$ used below in \Cref{sec:csbc}. This makes almost no difference in comparison to the case without the additional measure factor, where $m=0$ and $q^m$ stays at $1$. Therefore the additional measure factor does not affect much the integral over the original real contour. Yet as in the $N$ case, it generates a log singularity for the exponent, which sets $q(t_i)=0$ unchanged under the holomorphic gradient flow. Consequently, the ``side contour'' has no extension, and offers no contribution to the integral. This ensures that the integrals agree along the flowed contour and the original contour.

In summary, we will apply the generalized thimble method to the path integral
\begin{align}\label{eq:ZN1}
Z[q_0,q_1]=&\int_0^\infty \dd N \int_0^\infty dq(t_1)\cdots \int_0^\infty dq(t_{n}) ~ e^{i\sum_{i=0}^{n} \tilde{S}_i-\frac{n+1}{2} \log N},
\\
\tilde{S}_i=&2 \pi^2 \left(-\frac{3(q(t_{i+1})-q(t_{i}))^2}{4N \Delta t}+N \left(3k-\frac{1}{2} \Lambda  (q(t_{i})+q(t_{i+1}))\right)\Delta t\right)
\nonumber\\
&-\frac{i m}{2}\left(\log q(t_{i})+\log q(t_{i+1})\right),\label{eq:St}
\end{align}
where the measure factor for $q$ is absorbed in $\tilde{S}_i$, and that for $N$ is added to the exponent of the integrand. For convenience of writing, we separated a single factor $q(t_i)^m$ into two places in $\tilde{S}_i$ and $\tilde{S}_{i-1}$. This introduces constant factors for the unintegrated boundary $q$ values, but these constants drop out eventually when taking ratios as in \eqref{eq:expo}. The integrals in \eqref{eq:ZN1} are now for the Borel measure without additional factors, so \eqref{eq:fe} is directly applicable. 

\subsection{Notes on implementing the algorithm}

The generalized thimble method requires the integrand to be holomorphic everywhere the holomorphic gradient flow transverses. Since log functions show up in the integrand \eqref{eq:ZN1}, the integration domain is now taken on the Riemann surfaces of the $q$'s and $N$. This means in step 1 of the generalized thimble algorithm, one needs to keep track of the log branches for $q$ and $N$ during the flow. In the Julia programming language \cite{Bezanson2017Julia:Computing} that we use, this is implemented by the ``callback functions'' of the package ``DifferentialEquations.jl'' \cite{Rackauckas2022SciML/DifferentialEquations.jl:V7.3.0}, as is done in simplicial quantum gravity which refers to both log and square root branches \cite{Jia2022ComplexProspects}.

For step 2 of the generalized thimble algorithm, again as in \cite{Jia2022ComplexProspects} we apply the adaptive Monte Carlo sampler reviewed in \cite{Roberts2009ExamplesMCMC}. The rest of the steps are then implemented as stated in \Cref{sec:rm}.

\section{Lightcone fluctuations}\label{sec:lf}

The results of the generalized thimble method are in terms of the expectation values \eqref{eq:expo}. We are interested in $\ev{q(t)}$ and $\ev{N}$ for the squared scale factor and the lapse. 

In addition, we will compute the expectation values for the lightcone location and their fluctuations. Consider an event on the initial boundary for the minisuperspace universe at $t=0$, and set its coordinate to 
\begin{align}
(t,r,\theta,\phi)=(0,0,0,0)
\end{align}
in the spherical radial coordinate of \eqref{eq:metric}. We are interested in where its future lightcone cross the final boundary  for the minisuperspace universe at $t=1$. Since the metric \eqref{eq:metric} is spherical symmetric, in each metric the final lightcone location is the same for all $\theta,\phi$. Therefore we focus on the radial lightcone location. Since the path integral sums over different metrics, we will compute the expectation value for the final lightcone location.

For the metric \eqref{eq:metric}, the equation for radial lightlike geodesic is
\begin{align} \label{eq:ng}
	0=-\frac{N^2}{q(t)}\dd t^2+\frac{q(t)}{1-kr^2}\dd r^2.
\end{align}
With $\dd\chi^2=\frac{1}{1-kr^2}\dd r^2$, \eqref{eq:ng} implies
$\frac{N \dd t}{q(t)}=\dd \chi$. Integrating both sides yields
\begin{align}\label{eq:Dchi}
N \int \frac{1}{q(t)}\dd t=\Delta\chi.
\end{align}

During a time interval $\Delta t$, the zigzagging path of \eqref{eq:ZN} obeys $q(t)=\frac{q(t_{i+1})-q(t_{i})}{\Delta t}t+q(t_{i})$. Plugging this in \eqref{eq:Dchi} for yields
\begin{align}\label{eq:Dchi2}
\Delta\chi_i = \frac{N \Delta t (\log q(t_{i})-\log q(t_{i+1}))}{q(t_{i})-q(t_{i+1})}, \quad \Delta\chi=\sum_{i=0}^n \Delta\chi_i.
\end{align}
For a radial geodesic $\dd s^2=-\frac{N^2}{q(t)}\dd t^2+q(t)\dd \chi^2$, so $\Delta\chi$ gives the spatial comoving distance that a radial light ray covers from $t=0$ to $t=1$ and quantifies the size of the causal horizon for events at $t=0$. Below we will use \eqref{eq:Dchi2} in \eqref{eq:expo} to compute the expectation value $\ev{\Delta\chi}$ and the standard deviation
\begin{align}
    \sigma=\sqrt{\ev{\Delta\chi^2}-\ev{\Delta\chi}^2}
\end{align}
to quantify horizon fluctuations.

Such horizon fluctuations have been studied before in Lorentzian simplicial quantum gravity \cite{Jia2022LightGravity, JiaLightGravity} as an aspect where quantum cosmology may make a difference to standard paradigms of cosmology based on classical spacetimes. In particular, light rays that were never in causal contact if spacetime was treated classically could actually have been in causal contact if spacetime is treated quantumly to allow quantum fluctuations of the light ray paths. Therefore the horizon problem may take a different form in quantum cosmology in comparison to classical cosmology. Although it is too early to draw any definitive conclusions from the results presented below in \Cref{sec:csbc}, the potential for future developments should be bear in mind.

\section{Case study: bouncing cosmology}\label{sec:csbc}

In \Cref{fig:q-cosmology-cases_2} the bouncing saddle points $\bar{q}(t)$ are for $c_2=1$. When $k=1$, $\bar{q}(t)$ stays above $0$. When $k=0$, $\bar{q}(t)$ reaches $0$ at its minimum. When $k=-1$, $\bar{q}(t)$ drops below $0$. In the last two cases, the saddle point $\bar{q}(t)$ does not stay positive for all time, so $\bar{q}(t)$ is not included in the truly Lorentzian path integral sum. Therefore results from the real $q$ scheme run the risk of deviating much from the positive $q$ scheme. In this section we apply the method of \Cref{sec:gtm} to make a quantitative comparison between the positive and real $q$ schemes.

\subsection{Focusing on the bouncing saddle point}

As shown in \cite{Feldbrugge2017LorentzianCosmology}, both saddle points with $c_2=1$ and $c_2=-1$ in  \Cref{fig:q-cosmology-cases_2} are relevant for the real $q$ scheme path integrals. An otherwise unconstrained path integral will exhibit interference effects for the two saddle points. 

In comparing the positive and real $q$ schemes, we want to focus on the $c_2=1$ bouncing saddle point since the $c_2=-1$ saddle point stay high above $0$ for all time. One way to achieve this is to modify the boundary condition. The $c_2=1$ and $c_2=-1$ saddle points have different momentum at the boundaries. By employing a coherent state type boundary condition that centers around the momentum of $c_2=1$, one obtains a path integral without the interference of the other saddle point \cite{DiTucci2019QuantumInflation}.

The generalized thimble method offers a practical alternative. In a Monte Carlo simulation with multiple saddle points, the sampler needs to overcome the low integrand weight barrier at intermediate regions to move from around one saddle point to around another. Usually one wants to sample efficiently across different saddle points, and some advanced variations of the original generalized thimble method have been developed to achieve this \cite{Fukuma2017ParallelThimbles, Alexandru2017TemperedThimbles, Fukuma2021WorldvolumeMethod, Fujisawa2022BackpropagatingCalculations}. Here we want to avoid travelling across saddle points, which does not require any advanced method. Because of \eqref{eq:DreEDt}, a larger flow time $T$ increases the integrand weight barrier. Therefore we can simply adopt a large $T$ to restrict the Monte Carlo sampler to around the bouncing saddle point. The results presented next show the appropriate $T$ to achieve this.

\subsection{Results}\label{sec:r}

\begin{figure}
    \centering
    \includegraphics[width=1.0\textwidth]{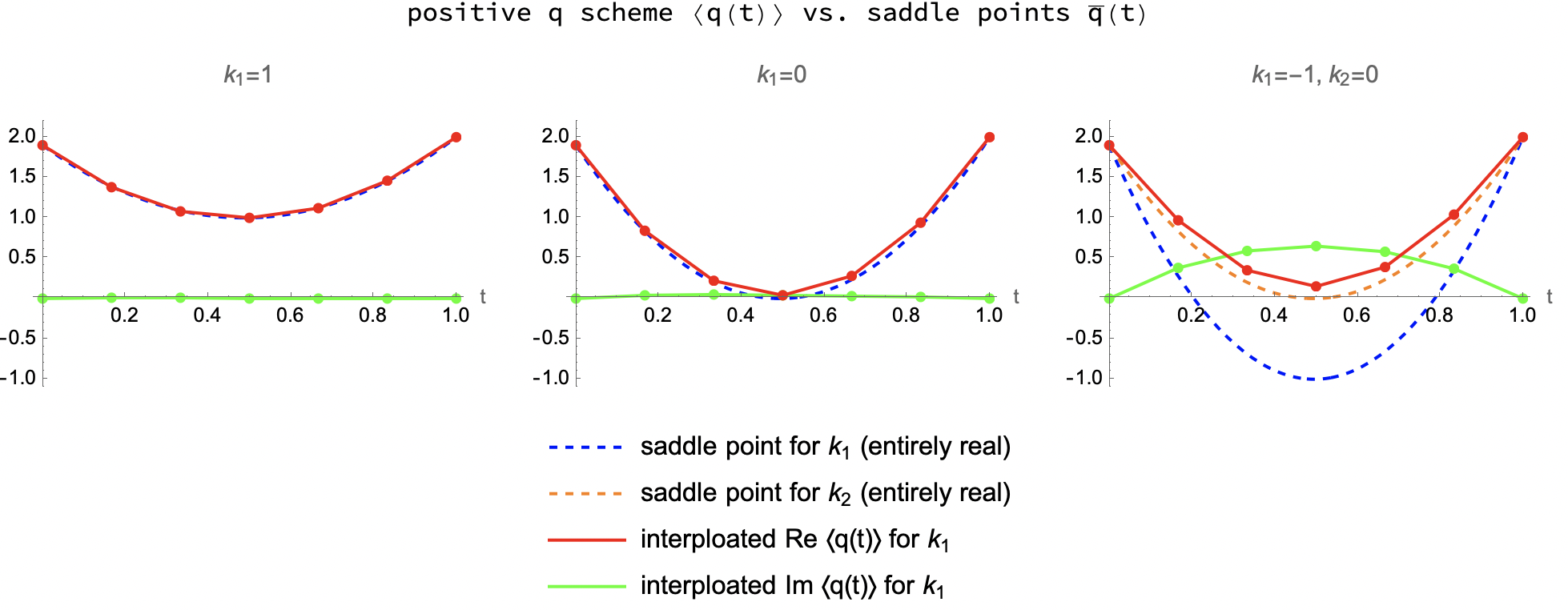}
    \caption{Comparing quantum expectation values  $\ev{q(t)}$ of \Cref{tb:r} from the positive $q$ scheme with the saddle point values $\bar{q}(t)$.}
    \label{fig:data-plot}
\end{figure} 

\begin{table}
    \centering
    \scalebox{0.7}{
    \begin{tabular}{c|cc|cc|cc}
    \hline
        $k$ & 1 & 1 & 0 & 0 & -1 & -1  \\ \hline
        scheme & positive & real & positive & real & positive & real  \\ \hline
        $m$ & -0.001 & 0 & -0.001 & 0 & -0.001 & 0  \\ \hline
        $T$ & 0.025 & 0.025 & 0.03 & 0.03 & 0.027 & 0.03  \\ \hline
        $\ev{e^{i\varphi}}$ & 1 & 1 & 0.98 & 0.98 & 0.94 & 0.96  \\ \hline
        $\ev{q(t_1)}$ &    1.38 + 0.01im &    1.41 - 0.01im &    0.84 + 0.04im &    0.86 + 0.02im &    0.97 + 0.38im &    0.22 - 0.01im  \\ \hline
        $\ev{q(t_2)}$ &    1.08 + 0.01im &     1.1 + 0.0im &    0.22 + 0.05im &    0.24 + 0.03im &    0.35 + 0.59im &   -0.77 - 0.02im  \\ \hline
        $\ev{q(t_3)}$ &     1.0 + 0.0im &     1.0 + 0.02im &    0.04 + 0.04im &    0.03 + 0.04im &    0.15 + 0.65im &    -1.1 - 0.02im  \\ \hline
        $\ev{q(t_4)}$ &    1.12 - 0.0im &     1.1 + 0.03im &    0.28 + 0.03im &    0.25 + 0.05im &    0.39 + 0.58im &   -0.74 - 0.01im  \\ \hline
        $\ev{q(t_5)}$ &    1.46 - 0.0im &    1.44 + 0.03im &    0.94 + 0.02im &    0.91 + 0.04im &    1.04 + 0.37im &    0.29 - 0.01im  \\ \hline
        $\ev{N}$ &    1.96 - 0.03im &    1.92 - 0.03im &     2.7 - 0.11im &    2.71 - 0.1im &    1.73 - 0.81im &     3.5 + 0.03im  \\ \hline
        $\ev{\Delta\chi}$ &   18.93 - 0.47im & N/A &   97.96 - 37.04im & N/A &     5.6 - 28.43im & N/A \\ \hline
        $\sigma$ &    1.08 - 0.08im & N/A &   11.84 + 18.21im & N/A &    1.07 - 1.9im & N/A \\ \hline
        $\abs{\sigma}$ & 1.08 & N/A & 21.72 & N/A & 2.18 & N/A \\ \hline
    \end{tabular}}
    \caption{Results for $q_0=1.9, q_1=2.0, \Lambda=3$. Data for each of the six columns is produced from a Monte Carlo chain of length $10$ million.}\label{tb:r}
\end{table}

In \Cref{fig:q-cosmology-cases_2} we took $q_0=1.9, q_1=2.0$ and $\Lambda=3$. For the same parameters, \Cref{fig:data-plot} and \Cref{tb:r} summarize the numerical results based on \eqref{eq:ZN1} with $n=5$.

For $k=1$ and $k=0$, the expectation values $\ev{q(t)}$ and $\ev{N}$ from the positive and real $q$ schemes are close, as seen in \Cref{tb:r}. The expectation values $\ev{q(t)}$ are in turn close to the saddle point values $\bar{q}(t)$, as seen from \Cref{fig:data-plot}. In particular, the imaginary part of the saddle points vanish, and the imaginary part of the expectation values are also close to zero. In contrast:
\begin{itemize}
\item For $k=-1$, the expectation values $\ev{q(t)}$ and $\ev{N}$ from the positive and real $q$ schemes differ much. In the positive $q$ scheme, $\Re\ev{q(t)}$ deviate much from $\bar{q}$ which is real, and $\Im\ev{q(t)}$ deviate much from zero.
\end{itemize}

Besides, there is a fundamental difference that applies to all values of $k$. In the positive $q$ scheme all path integral configurations are Lorentzian, so we can compute expectation values $\ev{\Delta\chi}$ for the comoving horizon \eqref{eq:Dchi2} and its fluctuations $\sigma$ as defined in \Cref{sec:lf}. In contrast, in the real $q$ scheme there are path integral configurations which are not Lorentzian, so the expectation value for the comoving horizon is undefined. 
The results for $\ev{\Delta\chi}, \sigma, \abs{\sigma}$ for the real $q$ scheme are shown in the last three columns of \Cref{tb:r}. To the extent that $\abs{\sigma}$ offers an indirect indicator of the amount of fluctuation:
\begin{itemize}
\item In the positive $q$ scheme, the amount of lightcone fluctuation as indicated by $\abs{\sigma}$ is much larger for $k=0$ than for $k=1$ and $k=-1$.
\end{itemize}

\subsection{Understanding the expectation values}

\begin{figure}
    \centering
    \includegraphics[width=0.5\textwidth]{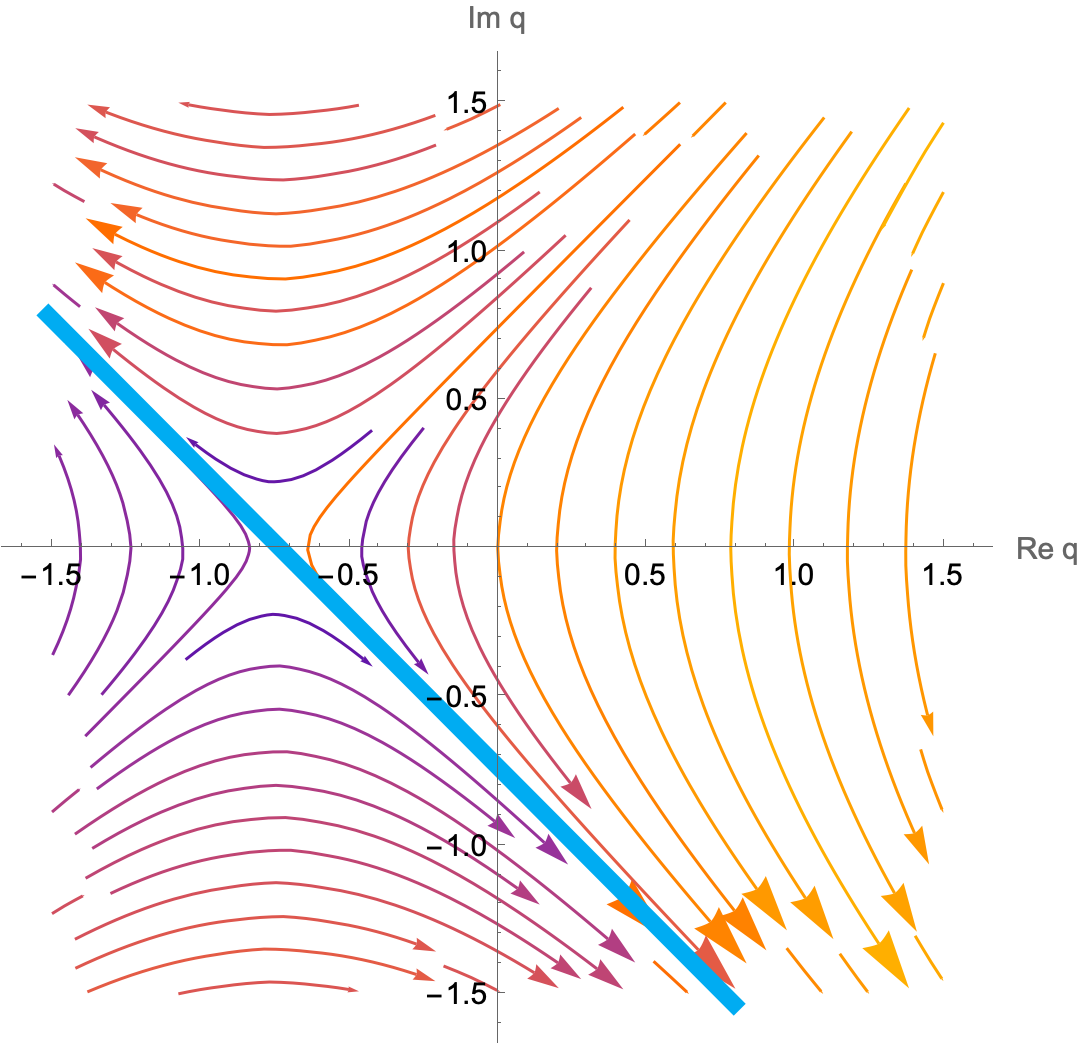}
    \caption{Holomorphic gradient flow for the one-variable model for $N=3$. The steepest descent contour is labelled by the thickened line.}
    \label{fig:hgf}
\end{figure} 

In the first bullet point of \Cref{sec:r}, we noted that the real $q$ scheme expectation values for $q$ differ much from the saddle point values in the negative spatial curvature $k=-1$ case. To understand this, it helps to consider again the one-variable toy model of \Cref{sec:tme}. According to Picard-Lefschetz theory \cite{Feldbrugge2017LorentzianCosmology}, the real line approaches the steepest descent contour asymptotically under the holomorphic gradient flow defined by \eqref{eq:piE} and plotted in \Cref{fig:hgf}. 

In the real $q$ scheme, the integration contour is the real line. Since this contour is deformed into the steepest descent contour under the flow, and the integral can be equivalently performed there. As shown in \Cref{fig:hgf}, the saddle point is at $-0.75$. Points around the saddle point along the steepest descent contour all have negative real parts around $-0.75$, but some have positive and some have negative imaginary parts. Therefore we expect $\ev{q(t)}$ to have a negative real part around the saddle point, and an almost vanishing or exactly vanishing imaginary part due to the cancellation from positive and negative contributions.

In the positive $q$ scheme, the integration contour is the positive halfline. As shown in \Cref{fig:hgf}, this contour only approaches a portion of the steepest descent contour quite far from the saddle point. In particular, all points of the original contour flow towards the directions of positive real values and negative imaginary values. Therefore we expect $\ev{q(t)}$ to have a much larger real part than the saddle point, and a much smaller imaginary part than zero. 

\begin{figure}
    \centering
    \includegraphics[width=0.8\textwidth]{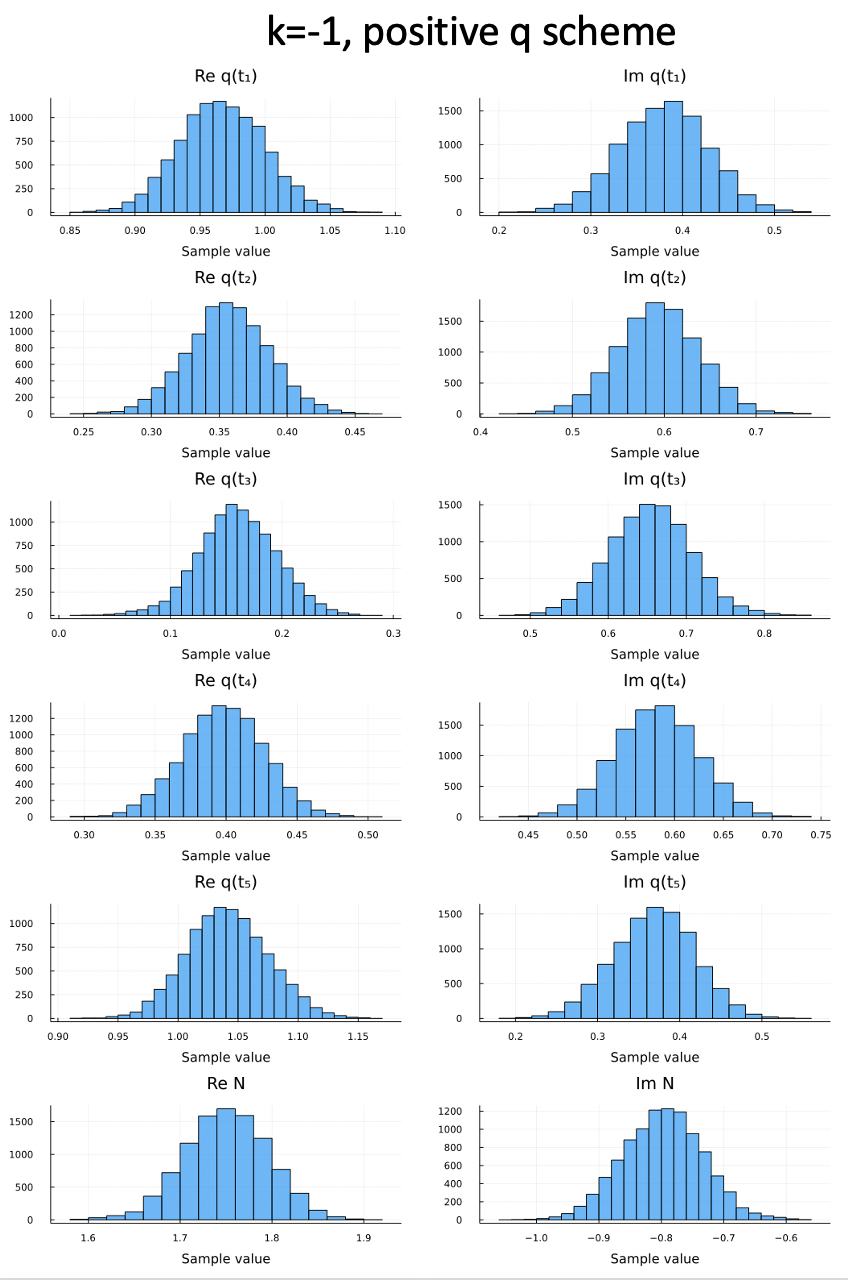}
    \caption{Histograms for the Monte Carlo sampling data for $k=-1$ in the positive $q$ scheme. To reduce complexity the length of the samples is reduced from 1000 million to 1 million by sequentially picking the first element from every 1000 samples.}
    \label{fig:qc-histogram3}
\end{figure}

The actual model for $k=-1$ we considered has more dynamical variables than one. However, similarly the real parts of $\ev{q(t)}$ deviate much from the negative saddle point values, and the imaginary parts of $\ev{q(t)}$ deviate much from zero (\Cref{fig:qc-histogram3}). Presumably this is for the same reason that under the holomorphic gradient flow, the original positive $q$ contour approaches only a portion of the steepest descent contour which does not cover the saddle point.


\begin{figure}
    \centering
    \includegraphics[width=0.8\textwidth]{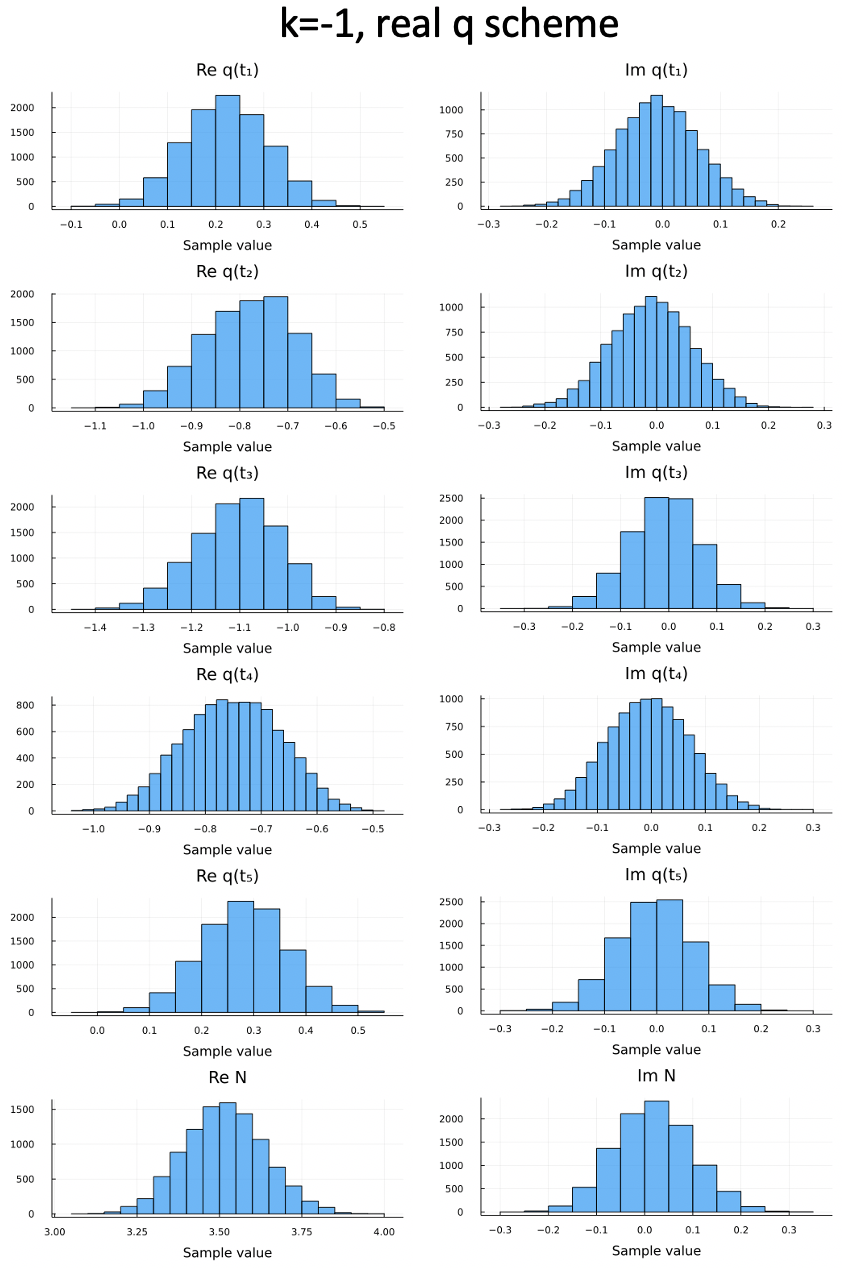}
    \caption{Histograms for the Monte Carlo sampling data for $k=-1$ in the real $q$ scheme. To reduce complexity the length of the samples is reduced from 1000 million to 1 million  by sequentially picking the first element from every 1000 samples.}
    \label{fig:qc-histogram4}
\end{figure} 

\subsection{Understanding the fluctuations}

From \Cref{tb:r}, we see that the lightcone fluctuation measurers $\sigma$ and $\abs{\sigma}$ differ much among the $k=1,0,-1$ cases. In addition, for $k=-1$ the fluctuations in $q$ and $N$ differ much between the real and positive q schemes, as shown in \Cref{fig:qc-histogram3} vs. \Cref{fig:qc-histogram4}. These difference can be understood better by drawing an analogy to a simpler setting.

\begin{figure}
    \centering
    \includegraphics[width=0.6\textwidth]{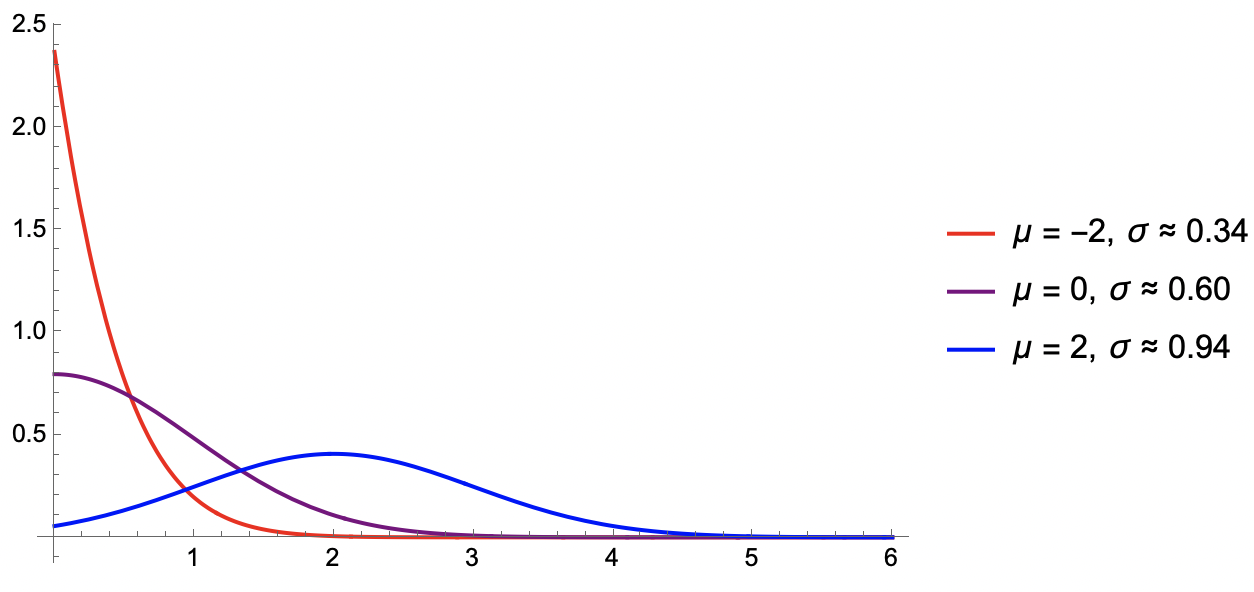}
    \caption{Sections of Gaussian distributions centered around $\mu$ and normalized over $\mathbb{R}^+$. The standard deviation $\sigma$ is smaller for smaller $\mu$.}
    \label{fig:qc-sdd}
\end{figure} 

Consider three Gaussian distributions with standard deviation $1$, but centered around different locations $\mu=-2,0$ and $2$. Assume that only the portion along the positive real halfline $\mathbb{R}^+$ is relevant, and rescale the distributions so that they are normalized on $\mathbb{R}^+$. As shown in \Cref{fig:qc-sdd}, the smaller $\mu$ is, the more sharply the distribution is peaked. 
As a consequence, the standard deviation $\sigma$ computed on $\mathbb{R}^+$ is smaller for smaller $\mu$ (\Cref{fig:qc-sdd}).

In the case of quantum cosmology, the real vs. positive $q$ schemes of $k=-1$ is like $\mu=2$ vs. $\mu=-2$. In the real $q$ scheme, the original contour covers the whole steepest descent contour under the holomorphic gradient flow, and the magnitude of the integrand varies slowly around the saddle point to yield relatively large fluctuations. This is like the $\mu=2$ case where a large portion around the peak $2$ is covered by the positive domain. In the positive $q$ scheme, the original contour only covers a portion of the steepest descent contour under the holomorphic gradient flow. This portion does not contain the saddle point, and the magnitude of the integrand varies much faster around the peak that is covered.  This is like the $\mu=-2$ case where the peak $-2$ lies far outside the positive domain. Therefore the fluctuations in $\ev{q(t)}$ and $\ev{N}$ are smaller in comparison to the real $q$ scheme (\Cref{fig:qc-histogram3} vs. \Cref{fig:qc-histogram4}). 

The positive $q$ scheme of $k=-1$ vs. $k=0$ is like $\mu=-2$ vs. $\mu=0$. For $k=0$, the saddle point is on the boundary of the original contour. This is like the $\mu=0$ case where the peak $0$ is on the boundary of the positive domain. Here we expect more fluctuation than $k=-1$, which is analogous to the $\mu=-2$ case where the peak $0$ lies far outside of the positive domain. This explains why there is more fluctuation for $k=0$ than for $k=-1$ measured by $\abs{\sigma}$, as noted in the second bullet point of \Cref{sec:r}.


Finally, there is also less fluctuation for $k=1$ than for $k=0$ as measured by $\abs{\sigma}$. This is presumably because of the form $\Delta\chi_i \propto \frac{\log q-\log q'}{q-q'}$ of the comoving distance function \eqref{eq:Dchi2}. As $q$ approaches zero, $\Delta\chi_i$ becomes very large and very sensitive to the precise value of $q$. Since in the $k=0$ case the saddle point value $q$ does get very close to zero, while in the $k=1$ case it does not, larger fluctuations in $\Delta\chi$ is expected.

\subsection{Breakdown of saddle point approximation}

In previous works of quantum cosmology, it is common to apply saddle point approximation to the path integrals. Here in the $k=-1$ example the saddle point failed at capturing the quantum expectation values of the truly Lorentzian path integral. This shows that the technique of saddle point semiclassical approximation does not enjoy universal validity for Lorentzian quantum cosmology, and must be used with caution.\footnote{When naive saddle point approximation on unrestricted domain does break down, it may be interesting to develop new saddle point approximation method on bounded domains along the line of, e.g., \cite{Delabaere2002GlobalBoundaries}.}


\section{Singularity avoidance}\label{sec:sa}

\subsection{Is singularity avoidance trivial?}

Even when the expectation values are close (as in the $k=0$ case), the positive and real $q$ schemes still differ on the critical issue of singularity avoidance. There are many non-trivial ideas of singularities avoidance in quantum gravity. For example, through discreteness, nonlocality, higher-order terms in the action, final boundary condition choices etc. However, there is a trivial alternative. A gravitational path integral may simply not include singular spacetimes in its sum \cite{JiaIsTrivial, Suen1989WaveSystem}.

For the minisuperspace model studied here, this indeed follows from including only Lorentzian configurations in the path integral. At the $q=0$ singularity, the metric \eqref{eq:metric} is of signature $(\infty,0,0,0)$. This is not of the Lorentzian signature, so it is automatically avoided in the truly Lorentzian path integral. In this sense, singularity avoidance is trivially achieved.

\subsection{Tunnelling and no-boundary proposals}

Interestingly, insisting on a strictly Lorentzian path integral for all time including $t=0$ invalidates from the outset Lorentzian variants of the tunnelling/no-boundary proposals that set $q(0)=0$. Therefore one must choose from: (1) allowing non-Lorentzian configurations in the path integral; or (2) rejecting boundary conditions that set $q$ to zero. 

Choice (1) calls for some additional specifications. Suppose quantum cosmology is governed by some fundamental theory of quantum gravity. Then how exactly are non-Lorentzian configurations included in the path integral for this fundamental theory? Is non-Lorentzianess only allowed at certain places of quantum spacetime but not others? If so, at exactly which kind of places, and why not at other places? 
One possibility is to consider non-Lorentzian pieces at the boundary of superspace, and allow this kind of non-Lorentzianness in the path integral \cite{Vilenkin1988QuantumUniverse}. However, this proposal needed to divide the boundary of superspace into regular and singular parts, and append additional rules associated with probability fluxes to these two parts. However, as far as we know the exact definition of the regular and singular parts of the boundary has never been written down in general \cite{Fanaras2022TheCosmology}, and this proposal still remains as an incomplete idea.

Choice (2) is dynamically less ambiguous since no additional rule is needed on how to include non-Lorentzian configurations. However, it leaves open the question of boundary conditions which can only be determined by other means. One possibility is to impose an ordinary Lorentzian boundary condition concentrated around small positive values of $q$. Another possibility is to give up on boundary conditions at small sizes of the universe and investigate boundary conditions for bouncing cosmology \cite{Brandenberger2017BouncingProblems}.

\section{Discussions}\label{sec:d}

Quantum cosmology based on Lorentzian path integrals is a promising avenue. However, many previous studies integrate the squared scale factor over the whole real line. This step introduces non-Lorentzian configurations into the path integral. Instead, a truly Lorentzian path integral should only include positive squared scale factor.

Here we studied and compared minisuperspace path integrals with real and positive squared scale factors. By restricting to Lorentzian configurations, the truly Lorentzian case enables the study of causal horizons and their quantum fluctuations, and achieves singularity avoidance by excluding singular minisuperspace geometries as non-Lorentzian. In addition, we find that the expectation values can differ much between the two cases. This happens in particular when the saddle point configuration does not belong to the strictly Lorentzian integration contour and is not connected to the strictly Lorentzian integration contour by the holomorphic gradient flow. 


These results challenge the universal validity of saddle point approximation widely used in quantum cosmology. In particular this affects topics such as Lorentzian variations of tunnelling/no-boundary proposals, and the quantum completeness of inflation \cite{DiTucci2019QuantumInflation}. In these cases, the saddle point gets close to or below zero, so that it does not belong to the strictly Lorentzian integration contour.
Instead of using saddle point approximation, a safer option is to compute the path integral directly. This can be done, for example, using the generalized thimble method adopted here.


We finish by a discussion on some topics to be understood better.

\subsection{Negative q}

Although the metric \eqref{eq:metric} has the $(-,+,+,+)$ signature only when $q>0$, it has the $(+,-,-,-)$ signature when $q<0$. One may wonder whether this rescues the real $q$ scheme for a Lorentzian path integral, since the $(+,-,-,-)$ signature might also be considered Lorentzian.

However, an attempt at rescue face some outstanding issues, because the real $q$ scheme path integral includes configurations where $q<0$ at certain times and $q>0$ at other times. 

First, such a configuration involves signature change. It does not qualify as a Lorentzian spacetime in the usual sense such that the spacetime stays within the $(-,+,+,+)$ or the $(+,-,-,-)$ signature. 

Second, in connecting the $q<0$ and $q>0$ parts of the configuration $q$ has to cross $0$. Here the metric has signature $(\infty,0,0,0)$. This is not Lorentzian.

Third, when $q$ crosses $0$, it is not \textit{a priori} clear what the causal structure is for that spacetime. Some additional rules are required to tell how causal paths travel across the singularity at $q=0$. Without such a rule, the causal relation between two events from the $q<0$ and $q>0$ parts of spacetime remains unclear.

\subsection{Inhomogeneity, anisotropy, and matter coupling}

The present study is restricted to minisuperspace models. For further research it is certainly interesting to accommodate inhomogeneity and/or anisotropy in the truly Lorentzian setting. 
For example, the Bianchi types I and III, and Kantowski-Sachs models studied in \cite{Halliwell1990Steepest-descentModels}, and the biaxial Bianchi IX model studied in \cite{Dorronsoro2018DampedState, Feldbrugge2018InconsistenciesProposal} may be simple enough as starting points to incorporate anisotropy. In a general non-perturbative setting, simplicial manifold models provide a systematic way to incorporate inhomogeneity and anisotropy in quantum cosmology \cite{Hartle1985SimplicialDiscussion, Hartle1986SimplicialTriangulations, Hartle1989SimplicalModel, Louko1992ReggeCosmology, Birmingham1995LensCosmology, Birmingham1998ACalculus, Furihata1996No-boundaryUniverse, Silva1999SimplicialField, Silva1999AnisotropicField, Silva2000SimplicialPhi2, daSilvaWormholesMinisuperspace}. Traditionally, simplicial quantum gravity is studied with respect to an Euclidean contour or an \textit{ad hoc} complex contour, but there has been growing attention towards the Lorentzian case \cite{Tate2011Fixed-topologyDomain, Tate2012Realizability1-simplex, Jia2022Time-spaceGravity, Jia2022ComplexProspects, Jia2022LightGravity, JiaLightGravity, JiaIsTrivial, Dittrich2022LorentzianSimplicial, AsanteComplexCosmology, Ito2022TensorCalculus}. In particular, the generalized thimble method employed here and in \cite{Jia2022ComplexProspects} may be applicable in studies of inhomogeneity and anisotropy. 

Certainly one should also consider matter coupling in further works. In addition to coupling to scalar fields and investigate the inflation scenario, we also find alternative scenarios without inflation worth investigating \cite{Gielen2015PerfectBounce, BoyleTwo-SheetedTime, TurokGravitationalPuzzles, BoyleThermodynamicConstant}.

\subsection{Lightcone topics}

Another topic about simplicial models of direct relevance is irregular lightcone structures. In simplicial models, there is the question whether the path integral should include simplicial geometries with interior points attached to more or fewer than two lightcones. In \cite{AsanteComplexCosmology, Dittrich2022LorentzianSimplicial} this question is studied based on a comparison with the continuum minisuperspace model in the real $q$ scheme. It is worth revisiting this topic given that the positive $q$ scheme may yield a different result.


In \Cref{sec:csbc}, we noted that the $k=0$ case exhibit larger causal horizon fluctuations than the $k=1$ and $k=-1$ cases. For the $k=-1$ case, that the fluctuations are smaller is related to the breakdown of saddle point approximation based on Einstein's equations. Whether this and other effects of lightcone fluctuations lead to any observable signatures is worth investigating further.

\subsection{Singularity}

Much of quantum cosmology is driven by the hope to understand singularities. In the recent wave of interest for Lorentzian quantum cosmology, the question has been raised whether singular geometries with $q=0$ should be avoided in the path integral \cite{DiTucci2019No-boundaryCosmology}, as such geometries enter critical discussions about boundary conditions \cite{Feldbrugge2017NoProposal, DiTucci2018UnstableMetrics, DiTucci2019No-BoundaryConditions, DiTucci2019No-boundaryCosmology, Jonas2022RevisitingField} and inflation \cite{Bramberger2019HomogeneousCosmology}. 
In \Cref{sec:sa} we showed that in a strictly Lorentzian path integral, singularities are automatically excluded as non-Lorentzian. How this affects the above topics is an open question.

There are many attempts to find effective regular spacetimes that replace spacetimes with cosmological and black hole singularities. Some of these derive regular solutions from equations of motion of modified actions. The $k=-1$ example studied here shows that an effective singularity-free geometry that characterizes the quantum theory at leading order (e.g., gives the correct expectation values) need not obey the equation of motion from an action principle. It remains to be clarified how such alternative views on singularity avoidance stand to each other. 

\subsection{Analytic insights}

In computing the oscillating complex path integrals, we applied the generalized thimble method \cite{Alexandru2016SignThimbles} to overcome the numerical sign problem. This method would not have been available a decade ago. However, new methods for evaluating complex path integrals are being developed at a promising pace in the recent decade (see e.g., \cite{Alexandru2022ComplexProblem, Berger2019ComplexPhysics, Gattringer2016ApproachesTheory, CarmenBauls2020ReviewTheories} and references therein). We expect such technical tools to boost the study beyond semiclassical analysis for the Lorentzian path integrals.

That said, it is still beneficial to find analytic methods to complement the numerical methods. One idea is to identify the value of $N$ so that the $N$-dependent saddle point $\bar{q}$ of \eqref{eq:qbar} just falls within the Lorentzian domain. Flowing this pair of $N$-$\bar{q}$ values under the holomorphic gradient flow to the steepest descent contour may yield a close guess at the expectation values.

\section*{Acknowledgement}

I am very grateful to Shengqi Sang, Latham Boyle, Niayesh Afshordi for helpful discussions, to Jean-Luc Lehners for friendly and informative comments on an earlier draft of the paper, and to Lucien Hardy and Achim Kempf for long-term encouragement and support. The numerical computation is conducted on Perimeter Institute's Symmetry high-performance computing cluster. I thank Dustin Lang for helps with setting up the computation there. Research at Perimeter Institute is supported in part by the Government of Canada through the Department of Innovation, Science and Economic Development Canada and by the Province of Ontario through the Ministry of Economic Development, Job Creation and Trade. 

\bibliographystyle{unsrt}
\bibliography{mendeley.bib}

\end{document}